\documentclass[12pt]{article}

\usepackage{amsmath}
\usepackage{graphicx}
\begin{document}

\hoffset = -0.7truecm
\voffset = -1.9truecm

\title{\bf
Monopole-Antimonopole Pair With Higher Energy Branch In  SU(2)×U(1) Weinberg-Salam Theory} 

\author{{Timothy Tie, Khai-Ming Wong, Dan Zhu}\\
\textit{{\small School of Physics, Universiti Sains Malaysia, 11800 USM, Penang, Malaysia}}}


\date{June 2021}
\maketitle

\begin{abstract}
We investigate the monopole-antimonopole pair solution in the SU(2) × U(1) Weinberg-Salam theory with $\phi$-winding number, $n=3$ for bifurcation phenomena. The magnetic monopole merges with antimonopole to form a vortex ring with finite diameter at $n=3$. Other than the fundamental solution, two new bifurcating solution branches were found when Higgs coupling constant $\lambda$, reaches a critical value $\lambda_c$. The two new branches possess higher energies than the fundamental solutions. These bifurcating solutions behave differently from the vortex ring configuration in SU(2) Yang-Mills-Higgs theory since thery are full vortex-ring. We investigate on the total energy $E$, vortex ring diameter $d_{\rho}$, and magnetic dipole moment $\mu_m$, for $0 \leq \lambda \leq 49$.
\end{abstract}


\section{Introduction}

The SU(2) Yang-Millls-Higgs (YMH) theory possesses a lot of interesting magnetic monopole solutions. One important solution is the spherically symmetric ‘t Hooft-Polyakov monopole solution with finite energy \cite{kn:1}. When $n$ ‘t Hooft-Polyakov monopoles superimpose at one location, a multimonopole, or $n$-monopole is formed \cite{kn:2}. Previous works in SU(2) YMH theory showed the existence of monopole-antimonopole pair (MAP), monopole-antimonopole chain (MAC) and vortex ring solutions \cite{kn:3}-\cite{kn:4}. Bifurcation of MAP, MAC and vortex ring configurations appears when $\phi$-winding number $n=3$. These branches with higher energies emerge when the Higgs coupling constant $\lambda$, reaches a critical value $\lambda_c$ .

Years ago, Y. Nambu predicted the existence of a monopole and antimonopole bound by flux string in the SU(2) × U(1) Weinberg-Salam model \cite{kn:5}. The configuration possesses finite energy and its mass is estimated in the TeV range. Ref. \cite{kn:6} discussed studies on sphaleron, a static saddle point particle-like solution, which was then identified to be consisting of a MAP and electromagnetic current loop. On the other hand, considering toplogically non-trivial sector, a spherical symmetric monopole solution is found to exist in Weinberg-Salam model with magnetic charge of $4 \pi / e$ and without a flux string \cite{kn:7}.

Recently we reported on the existence of axially symmetric MAP, MAC, and vortex-ring configuration with finite energy in SU(2) $\times$ U(1) Weinberg-Salam theory \cite{kn:8}. For the MAP system, the solution possesses zero net magnetic charge and a loop of electric current circulating it. While there is a finite separation between the two poles, it is found that a neutral flux string is connecting the monopole and antimonopole. This result is in line with Nambu’s work \cite{kn:5}. When $\phi$-winding number $n = 3$, vortex rings started to appear.

In this paper we investigate further the MAP configuration in the SU(2) $\times$ U(1) Weinberg-Salam theory for bifurcation phenomena when $n = 3$. We solved the equations of motion for $0 \leq \lambda \leq 49$ and Weinberg mixing angle, $\theta_W = 45^o$. Bifurcation started to appear at critical value of Higgs coupling constant $\lambda = \lambda_c$. Two new bifurcating solutions with finite but higher energies occurs, other than the fundamental solution. We compare these results with the bifurcating solution in SU(2) Yang-Mills-Higgs theory \cite{kn:4}.


\section{The Weinberg-Salam Model}

The Lagrangian in the standard Weinberg-Salam model is given by \cite{kn:7}
\begin{eqnarray}
&&{\cal L} = -{(\cal D}_\mu \boldsymbol{\phi})^\dagger ({\cal D}^\mu \boldsymbol{\phi}) - \frac{\lambda}{2}\left(\boldsymbol{\phi}^\dagger \boldsymbol{\phi} -\nu^2\right)^2 - \frac{1}{4}{\bf F}_{\mu\nu}\cdot {\bf F}^{\mu\nu} - \frac{1}{4}G_{\mu\nu}G^{\mu\nu},
\label{eq.1}\\
&&{\cal D}_\mu \boldsymbol{\phi} = \left(D_\mu - \frac{ig^\prime}{2} B_\mu \right) \boldsymbol{\phi}, ~~D_\mu = \partial_\mu - \frac{ig}{2} \boldsymbol{\sigma} \cdot {\bf A}_\mu.
\label{eq.2}
\end{eqnarray}
Here ${\cal D}_\mu $ is the covariant derivative of the SU(2) × U(1) group, $D_\mu$ is the covariant derivative for the SU(2) group. The gauge coupling constant, gauge potentials and electromagnetic fields of the SU(2) group are given by $g, {\bf A}_{\mu} = A^a_{\mu} \left(  \frac{\sigma^a}{2i}  \right)$ and $ {\bf F}_{\mu\nu} =  F^a_{\mu\nu} \left(  \frac{\sigma^a}{2i}  \right)  $ while the gauge coupling constant, gauge potentials and electromagnetic fields of the U(1) group are given by $g', B_{\mu}$, and $f_{\mu\nu}$ respectively. The term $\boldsymbol{\phi}$ is the complex scalar Higgs doublet, $\lambda$ is the Higgs field selfcoupling constant, the mass of the Higgs boson is $M_H = \nu \sqrt{2 \lambda}$, and $\nu$ is the Higgs field vacuum expectation value. The Higgs field can also be written as \cite{kn:7}
\begin{eqnarray}
\boldsymbol{\phi} = \frac{H(r,\theta)}{\sqrt{2}}   \boldsymbol{\xi}, ~~~\boldsymbol{\xi}^\dagger\boldsymbol{\xi}=1, ~~~
\hat{\Phi}^a = \boldsymbol{\xi}^\dagger\sigma^a \boldsymbol{\xi}, ~~~\sigma^a = \left(\begin{array}{ll}                   
																																												\delta^a_3 & \delta^a_1-i\delta^a_2\\
																																					 \delta^a_1+i\delta^a_2 & -\delta^a_3
																																					               \end{array}\right),
\label{eq.3}
\end{eqnarray}
where $\frac{H(r,\theta)}{\sqrt{2}}$ is the Higgs modulus, $\boldsymbol{\xi}$ is a column 2-vector, and $\hat{\Phi}^a$ is the Higgs unit vector.

The resultant equations of motion from Eq.(\ref{eq.1}) are
\begin{eqnarray}
&&{\cal D^\mu}{\cal D_\mu}\boldsymbol{\phi} = \lambda\left(\boldsymbol{\phi}^\dagger\boldsymbol{\phi}-\nu^2\right)\boldsymbol{\phi}, \nonumber\\
&&D^\mu {\bf F}_{\mu\nu} = -{\bf j}_\nu = \frac{ig}{2}\{\boldsymbol{\phi}^\dagger\boldsymbol{\sigma}({\cal D_\nu}\boldsymbol{\phi})-({\cal D_\nu}\boldsymbol{\phi})^\dagger\boldsymbol{\sigma\phi}\}, \nonumber\\
&&\partial^\mu G_{\mu\nu} = -k_\nu = \frac{ig^\prime}{2}\{\boldsymbol{\phi}^\dagger({\cal D_\nu}\boldsymbol{\phi})-({\cal D_\nu}\boldsymbol{\phi})^\dagger\boldsymbol{\phi}\}. 
\label{eq.4}
\end{eqnarray}
The energy density that corresponds to Eq.(\ref{eq.1}) is
\begin{eqnarray}
{\cal E} =\frac{1}{4}F^{a}_{ij}F^{a}_{ij} + \frac{1}{4}G_{ij}G_{ij} + {(\cal D}_\mu \boldsymbol{\phi})^\dagger ({\cal D}^\mu \boldsymbol{\phi}) + \frac{\lambda}{2}\left(\boldsymbol{\phi}^\dagger \boldsymbol{\phi} -\nu^2\right)^2 
\label{eq.5}
\end{eqnarray}


\section{The Magnetic Ansatz}

The electrically neutral axially symmetric magnetic ansatz \cite{kn:8} are
\begin{eqnarray}
&& g A^a_0 = 0,~~gA_i^a =  - \frac{1}{r}\psi_1~\hat{n}^{a}_\phi\hat{\theta}_i + \frac{n}{r}\psi_2~\hat{n}^{a}_\theta\hat{\phi}_i
+ \frac{1}{r}R_1~\hat{n}^{a}_\phi\hat{r}_i - \frac{n}{r}R_2~\hat{n}^{a}_r\hat{\phi}_i, \nonumber\\
&& \Phi^a = \Phi_1~\hat{n}^a_r + \Phi_2~\hat{n}^a_\theta = \Phi(r, \theta) \hat{\Phi}^a, 
\label{eq.6}\\
&& g^\prime B_0=0,~~g^\prime B_i = \frac{1}{r}B_1~\hat{\phi}_i, \nonumber\\
&& ~\boldsymbol{\xi} = i\left(\begin{array}{l}                   
																																												\sin\frac{\alpha(r,\theta)}{2}e^{-in\phi}\\
																																					 							 -\cos\frac{\alpha(r,\theta)}{2}
																																					               \end{array}\right), ~~
\hat{\Phi}^a = \boldsymbol{\xi}^\dagger\sigma^a \boldsymbol{\xi} = -\hat{h}^a, \nonumber																						\label{eq.7}														      
\end{eqnarray}
where $\psi_1, \psi_2, R_1, R_2, \Phi_1, \Phi_2, B_1$ are all functions of $r$ and $\theta$ and the unit vector $\hat{h}^a$ is given by \cite{kn:9}
\begin{eqnarray}
&& \hat{h}^a = h_1~\hat{n}^{a}_r + h_2~\hat{n}^{a}_\theta = \sin\alpha \cos n\phi~\delta^{a1} + \sin\alpha \sin n\phi~\delta^{a2} + \cos\alpha~\delta^{a3},  
\label{eq.7}\\
&& h_1 = \cos(\alpha-\theta), ~~h_2 =\sin(\alpha-\theta), ~~\alpha=\alpha(r,\theta). \nonumber
\end{eqnarray}
The angle $\alpha(r,\theta) \rightarrow p \theta$ when $r \rightarrow \infty$, where $p = 1, 2 ,3 ,...$ is an integer that indicates the number of magnetic poles \cite{kn:10}. Even value of $p$ result in MAP configuration. The spatial spherical coordinate unit vectors are 
\begin{eqnarray}
\hat{r}_i &=& \sin\theta ~\cos \phi ~\delta_{i1} + \sin\theta ~\sin \phi ~\delta_{i2} + \cos\theta~\delta_{i3}, \nonumber\\
\hat{\theta}_i &=& \cos\theta ~\cos \phi ~\delta_{i1} + \cos\theta ~\sin \phi ~\delta_{i2} - \sin\theta ~\delta_{i3}, \nonumber\\
\hat{\phi}_i &=& -\sin \phi ~\delta_{i1} + \cos \phi ~\delta_{i2}.
\label{eq.8}
\end{eqnarray}
and the isospin coordinate unit vectors are given by 
\begin{eqnarray}
\hat{u}_r^a &=& \sin \theta ~\cos n\phi ~\delta_{1}^a + \sin \theta ~\sin n\phi ~\delta_{2}^a + \cos \theta~\delta_{3}^a,\nonumber\\
\hat{u}_\theta^a &=& \cos \theta ~\cos n\phi ~\delta_{1}^a + \cos \theta ~\sin n\phi ~\delta_{2}^a - \sin \theta ~\delta_{3}^a,\nonumber\\
\hat{u}_\phi^a &=& -\sin n\phi ~\delta_{1}^a + \cos n\phi ~\delta_{2}^a,
\label{eq.9}
\end{eqnarray}

The magnetic ansatz is substituted into Eq.(\ref{eq.4}) and resulted in seven second order non-linear PDEs. They are
solved for all space for $0 \leq \lambda \leq 49$ with Weinberg angle, $\theta_W = 45^o$. The constants $g$ and $\nu$ are set to unity. The reduced equations are then solved by fixing boundary conditions at small and large $r$, as well as
along the $z$-axis and at $\theta = 0$ and $\theta= \pi$. The asymptotic solutions at large $r$ are \cite{kn:9}-\cite{kn:10}
\begin{eqnarray}
&& \psi_1(\infty,\theta) = 1,~~\psi_2(\infty,\theta) = 1+ \frac{\sin(\alpha-\theta)}{\sin\theta}(a \cos\theta+b), \nonumber\\
&& R_1(\infty,\theta) = 0,~~ R_2(\infty,\theta) = \cot\theta - \frac{\cos(\alpha-\theta)}{\sin\theta}(a \cos\theta+b), \nonumber\\
&& \Phi_1(\infty,\theta) = \nu \cos(\alpha-\theta),~~ \Phi_2(\infty,\theta) = \nu \sin(\alpha-\theta),
\label{eq.10}
\end{eqnarray}
while $B_1$ vanishes as for MAP. For MAP, $a=0$ and  $b=1$. The asymptotic solutions at small $r$ are the trivial solution,
\begin{eqnarray}
&&\psi_1(0,\theta) = \psi_2(0,\theta) = R_1(0,\theta) = R_2(0,\theta) = 0,\nonumber\\
&&\sin \theta~\Phi_1(0,\theta) + \cos \theta~\Phi_2(0,\theta)=0, \nonumber\\
&&\frac{\partial }{\partial r}(\cos\theta~\Phi_1(r,\theta)-\sin\theta~\Phi_2(r,\theta))|_{r=0}=0,
\label{eq.11}
\end{eqnarray}
and the boundary condition along the z-axis at $\theta=0$ and $\theta=\pi$ is
\begin{eqnarray}
\partial_{\theta} \psi_1 = \partial_{\theta} \psi_2 = R_1 = R_2 = \partial_{\theta} \Phi_1 = \partial_{\theta} \Phi_2 = B_1 = 0.
\label{eq.12}
\end{eqnarray}

\section{Electromagnetic Properties}
We choose to define the electromagnetic gauge potential and the neutral ${\cal Z}^0$ gauge potential by first gauge transforming the gauge potentials $A^a_\mu$ and Higgs field $\Phi^a$ of Eq. (\ref{eq.6}) to $A^{\prime a}_\mu$ and $\Phi^{\prime a}=\delta^a_3$ using the gauge transformation 

\begin{eqnarray}
&&U = -i\left[\begin{array}{ll}                   
																 \cos\frac{\alpha}{2}  				  & \sin\frac{\alpha}{2} e^{-in\phi}\\
																 \sin\frac{\alpha}{2} e^{in\phi} & -\cos\frac{\alpha}{2}
																 \end{array}\right]
= \cos\frac{\Theta}{2} + i\hat{u}_r^a \sigma^a \sin\frac{\Theta}{2}, 														 
\label{eq.13}\\
&&\Theta=-\pi ~~\mbox{and}~ ~\hat{u}_r^a = \sin\frac{\alpha}{2}\cos n\phi \delta^a_1 + \sin\frac{\alpha}{2}\sin n\phi \delta^a_2 + \cos\frac{\alpha}{2} \delta^a_3.	\nonumber	
\end{eqnarray} 
The transformed Higgs column unit vector and the SU(2) gauge potentials are 
\begin{eqnarray}
\xi^\prime &=& U\xi = \left[\begin{array}{l}                   
																 0\\
																 1
																 \end{array}\right]\nonumber\\
gA^{\prime a}_\mu &=& -gA^a_\mu - \frac{2}{r}\left\{\psi_2\sin\left(\theta-\frac{\alpha}{2}\right) + R_2\cos\left(\theta-\frac{\alpha}{2}\right)\right\}\hat{u}_r^a\,\hat{\phi}_\mu\nonumber\\
 &-& \partial_\mu\alpha\,\hat{u}_\phi^a - \frac{2n\sin\frac{\alpha}{2}}{r\sin\theta}\hat{u}_\theta^a\,\hat{\phi}_\mu.															 
\label{eq.14}
\end{eqnarray} 
or, in explicit form,
\begin{eqnarray}
gA^{\prime 1}_\mu &=& -\frac{1}{r}\cos n\phi \left\{\psi_2 h_1 + R_2 h_2 - \frac{n\sin\alpha}{\sin\theta}\right\}\hat{\phi}_\mu \nonumber\\
&-& \frac{1}{r}\sin n\phi\,(\psi_1-\partial_\theta \alpha)\,\hat{\theta}_\mu + \frac{1}{r}\sin n\phi\,(R_1+r\partial_r \alpha)\,\hat{r}_\mu								 
\label{eq.15}\\
gA^{\prime 2}_\mu &=& -\frac{1}{r}\sin n\phi \left\{\psi_2 h_1 + R_2 h_2 - \frac{n\sin\alpha}{\sin\theta}\right\}\hat{\phi}_\mu \nonumber\\
&+& \frac{1}{r}\cos n\phi\,(\psi_1-\partial_\theta \alpha)\,\hat{\theta}_\mu - \frac{1}{r}\cos n\phi\,(R_1+r\partial_r \alpha)\,\hat{r}_\mu								 
\label{eq.16}\\
gA^{\prime 3}_\mu &=& \frac{1}{r}\left\{\psi_2 h_2 - R_2 h_1 - \frac{n(1-\cos\alpha)}{\sin\theta}\right\}\hat{\phi}_\mu 
\label{eq.17}
\end{eqnarray}
Here we note that the gauge potential $gA^{\prime 3}_\mu$ (\ref{eq.17}) is actually the gauge potential that will give the 't Hooft electromagnetic field strength \cite{kn:1}, 
\begin{eqnarray}
\hat{F}_{\mu\nu} = \hat{\Phi}^a F^a_{\mu\nu} - \frac{1}{g}\epsilon^{abc}\hat{\Phi}^{a}D_{\mu}\hat{\Phi}^{b}D_{\nu}\hat{\Phi}^c 
= \partial_{\mu}A_\nu - \partial_{\nu}A_\mu - \frac{1}{g}\epsilon^{abc}\hat{\Phi}^{a}\partial_{\mu}\hat{\Phi}^{b}\partial_{\nu}\hat{\Phi}^c,
\label{eq.18} 
\end{eqnarray}
where $A_\mu = \hat{\Phi}^{a}A^a_\mu$ and $\hat{\Phi}^a = \Phi^a/|\Phi|$ as $\hat{\Phi}^{\prime a}=\delta^a_3$. Hence we denote the gauge potential $gA^{\prime 3}_\mu$ as the 't Hooft gauge potential.

The electromagnetic gauge potential ${\cal A}_\mu$ and the neutral $Z$-boson potential ${\cal Z}_\mu$ are defined as
\begin{eqnarray}
\left[\begin{array}{l}                   
			{\cal A}_\mu\\
			{\cal Z}_\mu
			\end{array}\right] = \left[\begin{array}{ll} 
			                      \cos\theta_W & \sin\theta_W \\
			                     -\sin\theta_W &  \cos\theta_W 
			                      \end{array}\right]         
\left[\begin{array}{l} 
B_\mu \\
A^{\prime 3}_\mu 
\end{array}\right]\nonumber\\\nonumber\\
= \frac{1}{\sqrt{g^2+g^{\prime 2}}}	\left[\begin{array}{ll} 
			                             g & g^\prime \\
			                     -g^\prime &  g
			                      \end{array}\right]         
\left[\begin{array}{l} 
B_\mu \\
A^{\prime 3}_\mu 
\end{array}\right]	                     			                   															 
\label{eq.19}
\end{eqnarray}	
where
\begin{eqnarray}
\cos\theta_W = \frac{g}{\sqrt{g^2+g'^2}} = \frac{e}{g'},~~\sin\theta_W = \frac{g'}{\sqrt{g^2+g'^2}} = \frac{e}{g},
\label{eq.20}
\end{eqnarray}
Hence the ${\cal A}_{\mu}$ potential and the neutral ${\cal Z}^0$ potential can be written as
\begin{eqnarray}
&& {\cal A}_{\mu} = \frac{1}{\sqrt{g^2+g'^2}} \left( g B_{\mu} + g' A^{3'}_{\mu} \right) = \frac{1}{e} \left( \cos^2\theta_W~g' B_{\mu} + \sin^2\theta_W~g A^{3'}_{\mu} \right), \nonumber\\
&& {\cal Z}_{\mu} = \frac{1}{\sqrt{g^2+g'^2}} \left( -g' B_{\mu} + g A^{3'}_{\mu} \right) = \frac{1}{e} \cos\theta_W \sin\theta_W \left( -g' B_{\mu} + g A^{3'}_{\mu} \right).
\label{eq.21}
\end{eqnarray}
where $gA^{3 \prime }_\mu$ is given in Eq.(\ref{eq.17}).

The U(1) magnetic field and the SU(2) 't Hooft magnetic field given by
\begin{eqnarray}
g^\prime B_i^{U(1)} &=& -\frac{g^\prime}{2}\epsilon^{ijk}G_{jk} = -\epsilon^{ijk}\partial_j \{B_1\sin\theta\}\partial_k\phi ~~\mbox{and}\nonumber\\
gB^{tHooft}_i&=& -\frac{g}{2}\epsilon^{ijk}\hat{F}_{jk} = -\epsilon^{ijk}\partial_j\{gA^{\prime 3}_k\}\nonumber\\
&=&-\epsilon^{ijk}\partial_j \{(\psi_2 h_2-R_2 h_1)\sin\theta - n(1-\cos\alpha)\}\partial_k\phi
\label{eq.22}
\end{eqnarray}
respectively can be shown by plotting for the respective magnetic field lines. The U(1) magnetic field lines can be shown by drawing the lines of constant \{$B_1\sin\theta$\} and the SU(2) t' Hooft magnetic field lines can be shown by drawing the lines of constant \{$(\psi_2 h_2-R_2 h_1)\sin\theta - n(1-\cos\alpha)$\}. The electromagnetic dipole moment (in the unit of $1/e$) is calculated by using the boundary condition at large $r$ at $\theta = \pi/2$,
\begin{eqnarray}
A_i \rightarrow \frac{1}{e} g' B_i = \left(  \frac{1}{er} \right) B_1~\hat{\phi}_i = - \frac{\mu_m \sin\theta}{r^2}~\hat{\phi}_i.
\label{eq.23}
\end{eqnarray}


Using finite difference approximation method, the seven reduced equations of motion are converted into a discretized system with non-equidistant grid of 90 × 80. Compactified coordinate $\tilde{x} = \frac{r}{r+1}$ is used covering regions $0 \leq \tilde{x} \leq 1$ and $0 \leq \theta  \leq \pi$. The equations are then solved numerically with MATLAB by selecting good initial guess.

\section{Numerical Results and Discussion}

\begin{figure}[tbh]
	\centering
	\hskip0in
	 \includegraphics[scale=0.23]{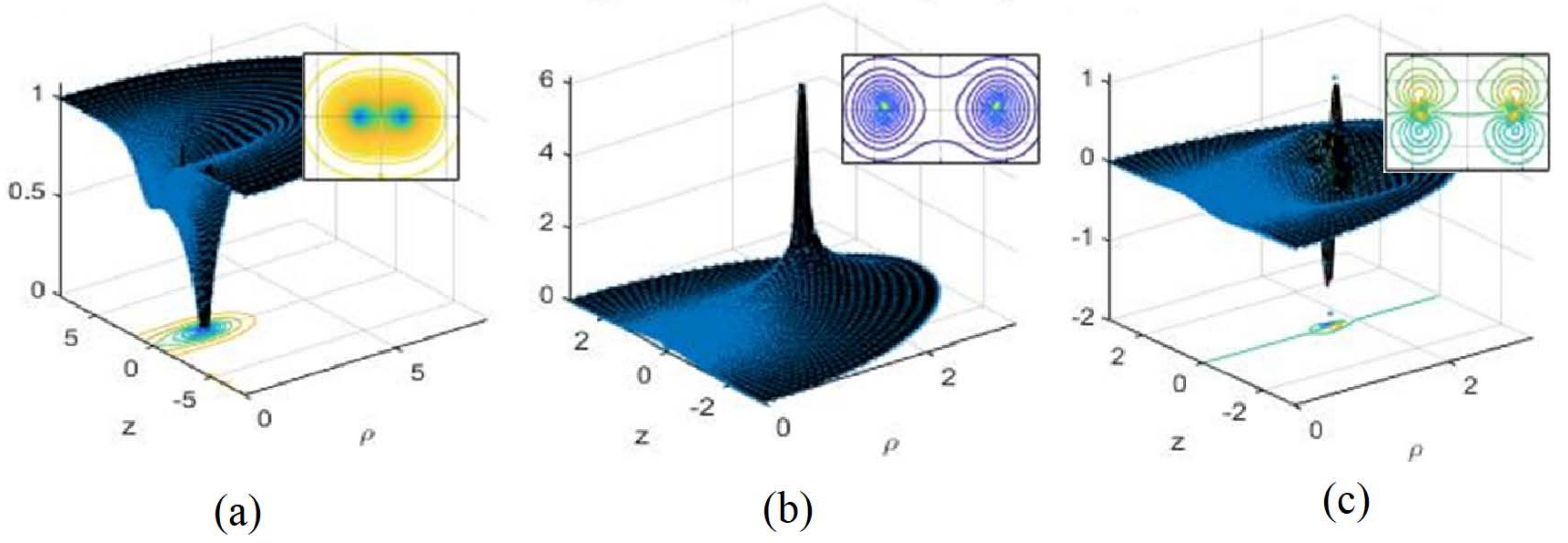} 
	\caption{Plots of (a) Higgs modulus, (b) energy density and (c) magnetic charge density of LEB when $n=3$ and $\lambda = 1$. }
 \label{Fig.1}
\end{figure}

Two new branches of vortex-ring configurations with higher energies than fundamental branch (FB) solution were found, namely Lower Energy Branch (LEB) and Higher Energy Branch (HEB). Figure 1(a) shows the Higgs modulus of the LEB solution, whereas Figures 1(b) and 1(c) show the energy density and magnetic charge density of the LEB solution when $\lambda = 1$. The results here exhibit some diferences from the work in Ref. \cite{kn:4}, where transition occurs between vortex ring and MAP configurations in the branches.  All the FB, LEB and HEB solutions here appear to be full vortex ring configurations within $0 \leq \lambda \leq 49$. We tabulate their values at certain $\lambda$ in Table 1.

From Figure 2(a), the total energy $\epsilon_n$ of the vortex ring increases with $\lambda$ for all branches, as expected. Bifurcation began at some critical point of $\lambda \approx 1$. Starting at $\lambda \approx 8$, the energies of LEB approaches energies of FB solution. Unlike the results in Ref. \cite{kn:4} where the energies of all branches saturate with increasing $\lambda$, the energies of solution branches obtained here does not show saturation within $0 \leq \lambda \leq 49$. However, it can be deduced from Figure 2(a) that the energy of FB solution will start to saturate at higher $\lambda$, but the energies of LEB and HEB solutions seemimgly will increase steadily.

\begin{figure}[tbh]
	\centering
	\hskip0in
	 \includegraphics[scale=0.21]{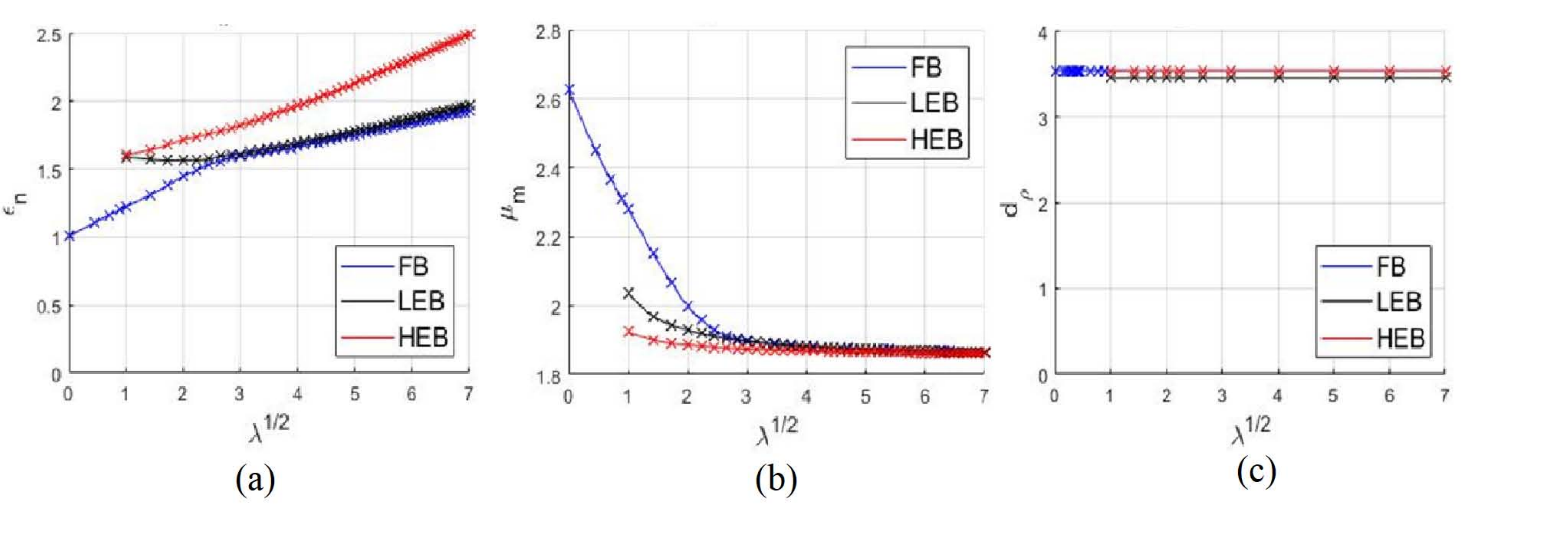} 
	\caption{Plots of (a) total energy, (b) magnetic moment, and (c) vortex ring diameter of FB, LEB and HEB versus $\sqrt{\lambda}$.}
 \label{Fig.2}
\end{figure}

The magnetic dipole moment $\mu_m$ of MAP decreases with increasing $\lambda$ and saturate gradually as shown in Figure 2(b). Similarly, the value of $\mu_m$ for LEB and HEB solutions approaches FB solution when $ \lambda \approx 8$. This phenomena however is again quite different from that in Ref. \cite{kn:4}, where the $\mu_m$ of bifurcating solutions has common value at $\lambda_c$. Here the $\mu_m$ of LEB and HEB solutions has different values at $\lambda = \lambda_c$. The vortex ring diameter $d_{\rho}$, is plotted against $\sqrt{\lambda}$ in Figure 2(c). For all FB, LEB and HEB solutions, their vortex ring diameters remain almost the same over the range of $\lambda$. This is slightly different from our previous finding \cite{kn:8}, most probably due to different resolution used. Figures 3(a) – 3(d) show the field lines of SU(2) field, U(1) field, electromagnetic field and neutral field for LEB solution when $\lambda = 1$. They obviously resemble the field lines reported in Ref. \cite{kn:8}.

\begin{figure}[tbh]
	\centering
	\hskip0in
	 \includegraphics[scale=0.36]{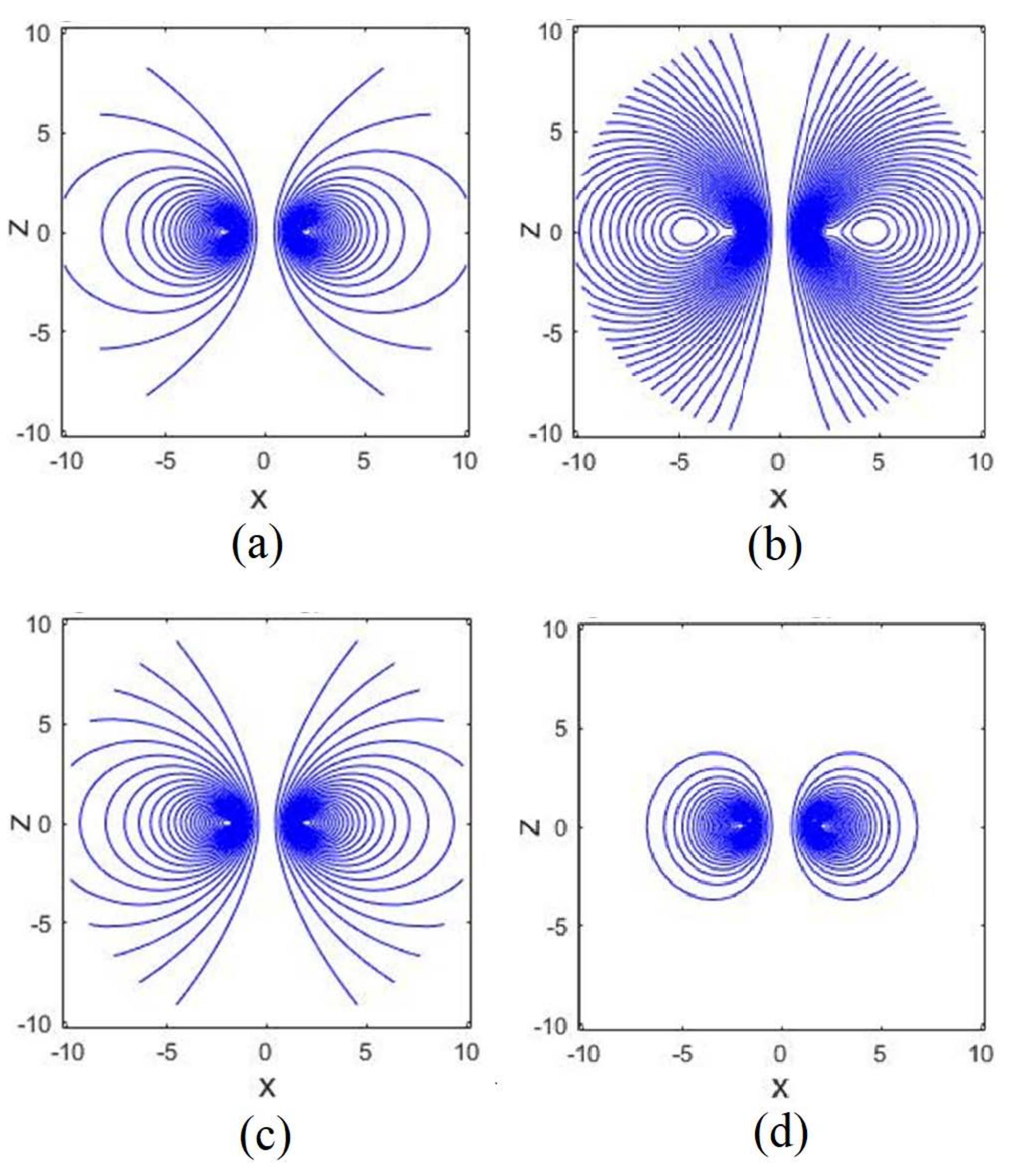} 
	\caption{Plots of countour field lines of (a) SU(2), (b) U(1), (c) electromagnetic and (d) neutral fields of LEB solutions at $\lambda = 1$. }
 \label{Fig.3}
\end{figure}

\begin{table}
\begin{tabular}{|c|c|c|c|c|c|c|c|c|c|}
\hline 
$\alpha$ & 0 & 1 & 4 & 9 & 16 & 25 & 36 & 49    \\ \hline
FB & 1.0058 & 1.2248 & 1.4475& 1.5942 & 1.6687 & 1.7517 & 1.8391 & 1.9297   \\ \hline
LEB & - & 1.5902 &  1.5664 & 1.6131 & 1.6893 & 1.7788 & 1.8733 & 1.9713        \\ \hline
HEB & -  & 1.6027 &  1.7161 & 1.8239 & 1.9674 & 2.1322 & 2.3099 & 2.4928      \\ \hline
\end{tabular}

\caption{Total energy ${\cal E}$ of FB, LEB and HEB solutions at various Higgs coupling constant $\lambda$ respectively.}
\label{table.1}
\end{table}

\section{Conclusions}

In conclusion, two new branches of bifurcating vortex-ring solutions were found in the SU(2) × U(1) Weinberg-Salam theory when $n = 3$, namely the LEB and HEB solutions. The energies of both branches are finite and higher than the fundamental solutions but they possess different behavior from their corresponding solutions in SU(2) Yang-Mills-Higgs theory. More work will be carried out along this line to clarify the origin of these differences. Future work will also be carried out to construct electrically charged MAP configurations in SU(2) × U(1) Weinberg-Salam theory.


\section*{Acknowlegdements}
The authors would like to thank School of Physics of Universiti Sains Malaysia, the Bridging Fund Grant (304/PFIZIK/6316278)

\newpage


\begin{thebibliography}{99}

\bibitem[1]{kn:1} G. ’t Hooft, Nucl. Phy. B 79 276 (1974); A.M. Polyakov, JETP Lett. 20 194 (1974).

\bibitem[2]{kn:2} C. Rebbi and P. Rossi, Phys. Rev. D 22, 2010 (1980); M.K. Prasad, Commun. Math. Phys. 80, 137 (1981); M.
K. Prasad and P. Rossi, Phys. Rev. D 24, 2182 (1981). 

\bibitem[3]{kn:3} B. Kleihaus and J. Kunz, Phys. Rev. D 61, 025003 (1999); B. Kleihaus, J. Kunz, and Y. Shnir, Phys. Lett. B
570, 237 (2003); Phys. Rev. D 68, 101701 (2003); Phys. Rev. D 70, 065010 (2004). 

\bibitem[4]{kn:4} J. Kunz, U. Neemann, and Y. Shnir, Phys. Lett. B 640, 57 (2006); R. Teh, A. Soltanian, K.M. Wong, Phys.
Rev. D 89, 045018 (2014); A. Soltanian, R. Teh, K.M. Wong, Int. J. Mod. Phys. A 31, Nos. 4 and 5, 1650006 (2016). 

\bibitem[5]{kn:5} Y. Nambu, Nucl. Phys. B 130, 505 (1977); Int. J. Theo. Phys. 17, 287 (1978). 

\bibitem[6]{kn:6} F.R. Klinkhamer, N.S. Manton, Phys. Rev. D 30 2212 (1984); M. Hindmarsh, M. James, Phys. Rev. D 49 6109
(1994); E. Radu, M.S. Volkov, Phys. Rev. D 79 065021 (2009). 

\bibitem[7]{kn:7} Y.M. Cho, D. Maison, Phys. Lett. B 391, 360 (1997); Kyoungtae Kimm, J.H. Yoon, Y.M. Cho, Euro. Phys. J. C 75: 67 (2015); Kyoungtae Kimm, J. H. Yoon, S. H.
Oh, Y. M. Cho, Mod. Phys. Lett. A 31, 1650053 (2016). 

\bibitem[8]{kn:8} R. Teh, B.L. Ng, K.M. Wong, Ann. of Phys. 362, 170 (2015). 

\bibitem[9]{kn:9} K.G. Lim, Rosy Teh, K.M. Wong, J. Phys. G: Nucl. Part. Phys. 39, 025002 (2012). 

\bibitem[10]{kn:10} Rosy Teh, P.Y. Tan, K.M. Wong, Int. J. Mod. Phys. A 27 (2012) 1250148. 








\end{thebibliography}
\end{document}